\documentclass[final] {aipproc}

\layoutstyle{6x9}

\begin{document}

\title{A combinatorial approach to discrete geometry}
\author{L. Bombelli}{
  address={University of Mississipi. USA}
}
\author{Miguel Lorente}{
  address={Universidad de Oviedo, Spain}
}
\begin{abstract}
We present a paralell approach to discrete geometry: the first one introduces Voronoi 
cell complexes from statistical tessellations in order to know the mean scalar curvature
in term of the mean number of edges of a cell [1]. The second one gives the restriction of
a graph from a regular tessellation in order to calculate the curvature from pure
combinatorial properties of the graph [2].

Our proposal is based in some epistemological pressupositions: the macroscopic continuous
geometry is only a fiction, very usefull for describing phenomena at certain sacales, but it
is only an approximation to the true geometry. In the discrete geometry one starts from a
set of elements and the relation among them without presuposing space and time as a background.
\end{abstract}
\maketitle
\noindent {\bf 1. Introduction. }
In recent years some approaches to quantum gravity have suggested the hypothesis of a
discrete space time [3] as a consequence of the combinatorial properties of spin networks
underlying the structure of space [4] and implemented with the hypothesis of
causal sets [5][6].

In our model we have to choose the discrete quantities in such a way that in the continuous
limit they become the classical ones. The direct
way consists on calculating some continuous quantity in a 2d-manifold. By the inverse way, we can start directly from the graph and find some embedding where the
corresponding quantities become analog, like the genus of some  graph [7], or the
curvature in a triangulated manifold. 

According to Bombelli one can scatter points in a
Lorentzian manifold, and then keep the statistical distribution of
points from which some discrete quantities can be defined, such as curvature, from
combinatorial properties of the set of relations [8].

\medskip
\noindent {\bf 2. Reflection groups and tessellations. }
Let $P$ a finite sided n-dimensional convex polyhedron in a metric space $X$, all of whose dihedral angles
are submultiple of $\pi$. Then the group generated by the reflection of $X$ in the sides of $P$, $\{ S_i \} $
is a discrete reflection group $\Gamma$ with respect to the polyhedron $P$.

Let $\Delta$ be an n-simplex in $X$ all of whose dihedral angles are submultiple of $\pi$.
The group $\Gamma$ generated by the reflections of X in the sides of
$\Delta$ is an n-simplex discrete reflection group. Notice that $X$ can be $S^n,E^n\ {\rm
or}\  H^n$. The classification of all the irreducible n-simplex (spherical, euclidean and hyperbolic)
reflection groups is complete [9].

Assume that $n=2$. Then $\Delta$ is a triangle in $X$, whose angles ${\pi  \over l},{\pi 
\over m},{\pi  \over n}$ are submultiple of $\pi$. If we call $T\left( {l,m,n} \right)$ the
group $\Gamma$ generated by the reflections in the sides of $\Delta, T\left( {l,m,n} \right)$
is call a triangle reflection group.

If $X=S^2$ the only spherical triangle reflection groups are:
$$T(2,2,2), \qquad T(2,2,n)\; n>2, \qquad T(2,3,3),  \qquad T(2,3,4), \qquad T(2,3,5)$$
If $X=E^2$ we have the euclidean triangle reflection groups:
$$T(3,3,3), \qquad T(2,4,4), \qquad T(2,3,6)$$
If $X=H^2$ we have the hyperbolic triangle reflection groups:
$$T(2,m,n) \; m\geq n\geq 3, \qquad T(l,m,n) \; l\geq m\geq n\geq 3$$
Geometrically a reflection can be represented by a linear transformation which fixes an
hyperplane pointwise and sends some non zero vector to its negative. In the metric space
$X$ we construct vectors $\left\{ {\alpha _i} \right\}$ in one to one correspondence to the
sides $\left\{ {S_i} \right\}$ defined before, in such a way that the angle between ${\alpha
_i}$ and ${\alpha _j}$ will be compatible with the values of $k_{ij}$, namely, $\vartheta
\left( {\alpha _i,\alpha _j} \right)={\pi  \over {k_{ij}}}$, ($k_{ij},$ positive integer o
infinite, $k_{ii}=1$).

In order to construct a reflection with respect to these vectors $\left\{ {\alpha _i}
\right\}$ we define a non-degenerate symmetric bilinear form on $X$ by the formulas
\begin{eqnarray*}
\left\langle {\alpha _i,\alpha _j} \right\rangle =-\cos {\pi  \over {k_{ij}}}
\end{eqnarray*} This expresion is interpreted to be $-1$ for $k_{ij}=\infty $. Obviously
$\left\langle {\alpha _i,\alpha _i} \right\rangle =1$, and $\left\langle {\alpha _i,\alpha
_j} \right\rangle \le 0$ for $i\ne j$. For each vector ${\alpha _i}$ we can define a
reflection ${S_i}$ on $X$:
$$S_i\beta =\beta -2\left\langle {\alpha _i,\beta } \right\rangle \alpha _i\quad ,\quad
\beta \in X$$ 
clearly $S_i\alpha _i=-\alpha _i$ and all $\gamma $ satisfying
$\left\langle {\alpha _i,\gamma } \right\rangle =0$ belong to a plane invariant under $S_i$.

It can be proved that the collection of the polyhedra obtained by the reflections on the
side of $\Delta $ is a tessellation of $X$  all the n-simplex (compact or non-compact)
reflection groups lead to regular tessellation of
$X$.

In Figure 1 we give now some examples of tessellations in $H^2$ generating by reflecting
in the sides of the hyperbolic triangle $T(2,3,8)$.

\medskip
\noindent{\bf 3. Gauss curvature of continuous tessellations}. Two dimensional tessellations in $X\left( {=S^2,E^2\,{\rm or}\,H^2} \right)$ are generated
by 2-simplex (triangle) reflection group. 

In $S^2$ the geodesic triangle are
spherical.

The excess of the interior angles of a spherical triangle is:
$$
\epsilon  = \alpha  + \beta  + \gamma  - \pi  = {\pi  \over l} + {\pi  \over m} + {\pi 
\over n} - \pi 
$$
It can be proved that this excess is always positive [9]

The area of the triangle $T\left( {x,y,z} \right)$ is 
\begin{eqnarray*}  {\rm Area} \left\{ {T\left( {x,y,z} \right)} \right\}=\alpha +\beta
+\gamma -\pi =\epsilon
\end{eqnarray*} 
The excess of the interior angles of an euclidean triangle is 
\begin{eqnarray*} 
\epsilon =\alpha +\beta +\gamma -\pi =0
\end{eqnarray*} In $H^2$ the geodesic triangles are hyperbolic.

The excess of the interior
angles of an hyperbolic triangle is
$$
\epsilon  = \alpha  + \beta  + \gamma  - \pi  = {\pi  \over l} + {\pi  \over m} + {\pi 
\over n} - \pi 
$$
It can be proved that this excess is always negative [9]

The area of an hyperbolic triangle $T\left( {x,y,z} \right)$ is
\begin{eqnarray*}  {\rm Area}\;(T)=\pi -\left( {\alpha +\beta +\gamma } \right)
\end{eqnarray*}
We can now apply these results to the curvature of the surfaces corresponding to the
2-dimensional regular tessellations (spherical, euclidean or hyperbolic). According to
Gauss-Bonet theorem the excess angle of some geodesic triangle $T$ is equal to the
integral of the gaussian curvature $K$ over $T$
\begin{eqnarray*}
\epsilon =\alpha +\beta +\gamma -\pi =\int\!\!\!\int\limits_T {Kd\sigma }
\end{eqnarray*} where $d\sigma$ is the area element. If $K$=const.
\begin{eqnarray*} K={\epsilon  \over A}
\end{eqnarray*}

Applying this formula to the above results, we have:

\qquad $K=1$ for spherical geodesic triangles

\qquad $K=0$ for euclidean triangles

\qquad $K=-1$ for hyperbolic geodesic triangle

\bigskip
\noindent{\bf 4. Curvature on planar graphs.} 
A graph is a par $G=\left\{ {V,E} \right\}$ where $V$ is a non-empty set of vertices and $E$
an unordered 2-set of vertices, called edges, in such a way that two vertices are incident
to an edge. 

A graph can be defined in an abstract way using only combinatorial properties of vertices
and edges, or can be obtained from geometrical objects. For instance, given a particulr
tessellation described in section 2, we keep the edges and vertices of all the triangles and
eliminate the embedding manifold in such a way that we are left with the points (vertices) and
relations among those (edges). In Figure 2, we have drawn the graphs that we have derived by
this method from the tessellations given in Figure 1, where the vertices are
represented by points and the edges by arrows.
\begin{figure}
\centering
\includegraphics[height=6cm]{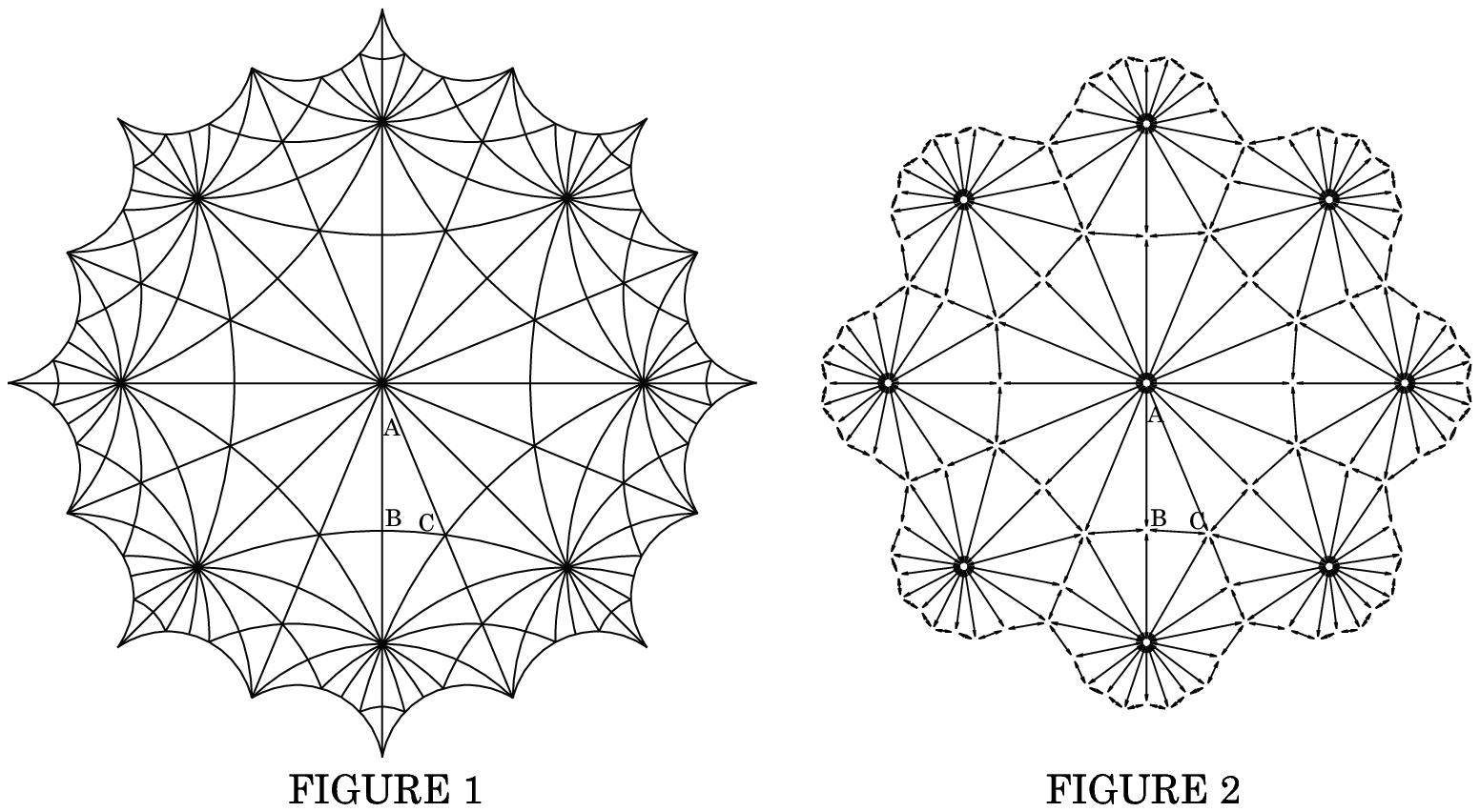}
\label{fig:6}       % Give a unique label
\end{figure}
In a graph one can define such elements as path, circuit, length, distance. For instance, in a
given graph one may travel from one vertex to another using several edges. the set of the vertices
visited in that journey is called a path. These definitions coincide with the
standard ones when the graph is embedded in some continuous manifold.

One can define the excess of
this triad of vertices as the quantity, in analogy with (1), 
$$\delta ={1 \over l}+{1 \over m}+{1 \over n}-1$$ where $2l, 2m, 2n$ are the number of edges
incident in each of the three vertices, which correspond to $2l$--valued, $2m$--valued or
$2n$--valued vertices, respectively.

If we define the spherical, euclidean or hyperbolic graph, that is obtained from a
spherical, euclidean or hyperbolic tessellation respectively, we can check

\qquad $\delta >0$, for a spherical graph

\qquad $\delta =0$, for an euclidean graph

\qquad $\delta <0$, for an hyperbolic graph

We define the area  and the area of the triad $T(l, m, n)$
in an hyperbolic graph 
$$\sigma (T)=1-\left( {{1 \over l}+{1 \over m}+{1 \over n}} \right)$$
Similarly, we define the curvature of a triad $T(l, m, n)$
$$K(T)={\delta  \over \sigma }=\left\{ \begin{array}{lr} {1,} &{\rm for\ a \ spherical\
graph} \\ {0,} &{\rm for\ an\ euclidean\ graph} \\ {-1,} &{\rm for\ an\ hyperbolic\ graph}
\end{array} \right.$$ 
an expression that can be considered the discrete version of the Gauss-Bonet theorem. 

\bigskip
\noindent{\bf 5. Statistical approach to Gaussian curvature. }
In a random distribution of points we can apply the combinatorial approach to calculate the
curvature as in the regular case.
\setcounter{figure}{2}
\begin{figure}
\centering
\includegraphics[height=6cm]{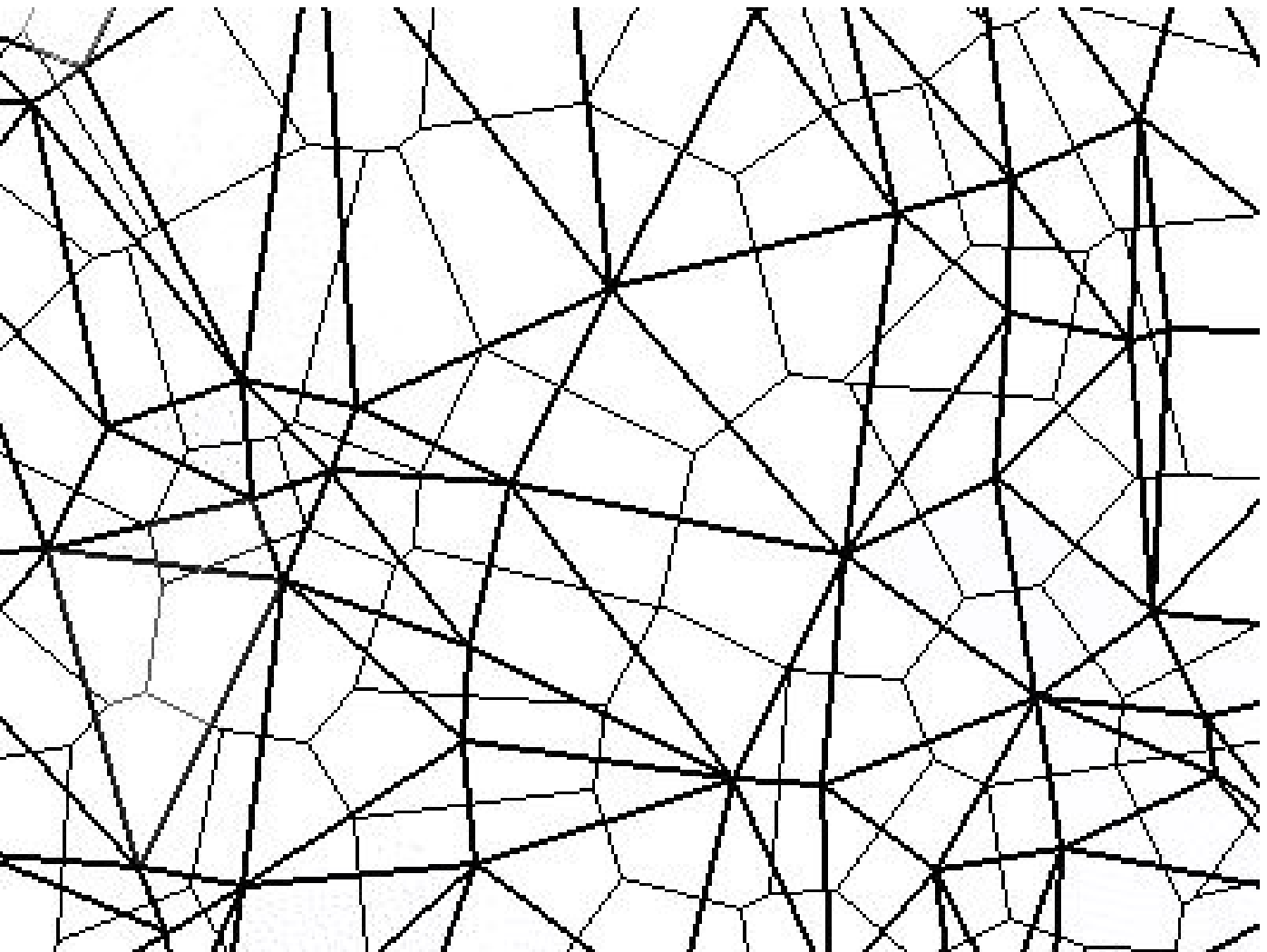}
\caption{Example of 2D Voronoi (thin lines) and Delaunay (thick lines) complexes; the
points they are based on are the Delaunay vertices.}
\end{figure}
We can consider a cell decomposition of a manifold by the embedding of a cell complex $\Omega$,
formed by $N_0$ vertices, $N_1$ edges, $N_2$ faces, $N_3$ tetrahedra and so on, satisfying
$$
\sum\limits_{k = 0}^D {\left( { - 1} \right)^K N_k \left( \Omega  \right)}  = \left( { - 1} \right)^D \chi \left( \Omega  \right)
$$
One of the most useful cell decomposition is a triangulation in which all the cell complex are
simplicial, satisfying $N_1 \left( \Omega  \right) = {1 \over 2}\left( {D + 1} \right)N_0 \left(
\Omega  \right)$. Given a random distribution of points $p_i$ we construct the Voronoi complex as
the set of points $p_j$ that are closer to $p_i$ than other point $p_j$. The Delaunay complex is the
dual to Voronoi, and it is formed by all the points $p_i$ and the edges joining them (see Figure 3).
A Voronoi complex gives rise to a non-regular tessellation. If we substract the embedding manifold
we are left with a graph, where we can calculate curvature from combinatorial properties of Voronoi
complex. Suppose we have a 2-dimensional Voronoi graph $\Omega $ with elements $N_0, N_1, N_2$
satisfying $N_1  = {1 \over 2}\left( {2 + 1} \right)N_0 \quad ,\quad N_0  - N_1  + N_2  = \chi \left( \Omega 
\right),\chi $\quad Euler number.

In order to calculate the mean curvature, we use the Gauss-Bonet formula:
$$
\bar N_1  = 2 \cdot {{N_1 \left( \Omega  \right)} \over {N_2 \left( \Omega  \right)}} = G\left( {1 - {{\chi \left( \Omega  \right)} \over {N_2 }}} \right) = G\left( {1 - {{\int {Kd\sigma } } \over {4\pi \rho A}}} \right) = G\left( {1 - {K \over {4\pi \rho }}} \right)
$$
whence $K = 4\pi \rho \left( {1 - {1 \over G}\bar N_1 } \right)$, where $\rho$ is the density of the cell complex in $\Omega$

\bigskip
This work was partially supported by Ministerio Educaci\'on y Ciencia, grant BFM 2003-00313/FIS

\end{document}